\documentclass[aps,prb,showpacs,floatfix,amsmath,amssymb,
               reprint]{revtex4-1}
\usepackage{color}
\usepackage{graphicx}
\usepackage{natbib}
\usepackage{epsfig}
\usepackage{setspace}
\usepackage{amsmath}
\usepackage{amssymb}
\usepackage{verbatim}
\usepackage{bibentry}
\usepackage{indentfirst}
\usepackage{titlesec}
\titlespacing*{\section}
{0pt}{4.5ex plus 1ex minus .2ex}{4.3ex plus .2ex}
\titlespacing*{\subsection}
{0pt}{1.5ex plus 0.5ex minus .2ex}{4.3ex plus .2ex}

\begin{document}
\title{Scaling analysis of the extended single impurity Anderson model: Renormalization due to valence fluctuations}
\author{Rukhsan Ul Haq }
\author{ N.\ S.\ Vidhyadhiraja}
\affiliation{Theoretical Sciences Unit,\\ Jawaharlal Nehru Centre for Advanced Scientific Research,\\ Bangalore, India}
\begin{abstract}
In this paper we have explored the role of valence fluctuations in an extended Anderson impurity model (e-SIAM) in which there is an additional Hubbard repulsion between conduction  and impurity electrons, employing perturbative renormalization methods. We have calculated the scaling equations for the model parameters and solved them both analytically and numerically to find how valence fluctuations renormalize these parameters. Analytical solutions of the scaling equations yielded scaling invariants of the model which we find to be different than the Anderson impurity model signifying different kind of scaling trajectories of e-SIAM. The strong coupling physics of SIAM is known to be governed by spin fluctuations and we hence analysed how the strong coupling regime of e-SIAM is renormalized by valence fluctuations. Doing a third order perturbative renormalization of the model gave us access to the Kondo scale. We not only confirmed that valence fluctuations enhance the Kondo scale but we also calculated the functional form of this dependence. Benchmarking some of our results with numerical renormalization group calculations, we found excellent agreement. Applying Schrieffer-Wolff transformation, we found that the strong coupling regime of the model is governed by spin-charge Kondo model unlike the Anderson impurity model. Our results suggest that spin Kondo effect can co-exist with valence fluctuation mediated charge Kondo effect.
\end{abstract}

\maketitle
\section{Introduction} 
The Doniach phase diagram has been employed as a generic framework to understand the physics of heavy fermion systems~\cite{book1,book2}, where spin fluctuations govern the low energy Kondo physics~\cite{Coleman}. Though one can understand a broad range of phenomenology in rare-earths in terms of 
spin fluctuations alone, there are many experimental 
observations such as first order valence transition, unconventional superconductivity in $CeCu_{2}Se_{2}$ and quantum criticality in  $YbRh_{2}Si_{2}$,
$\beta-YbAlB_{4}$ and $CeIrIn_{5}$
~\cite{Miyake1,Watanabe2}, that require incorporating valence fluctuations on an equal footing.
Historically the Falicov-Kimball (FK) model was introduced to investigate valence fluctuations. But, the FK model has spinless electrons; hence to get a 
realistic description, the Anderson model is a more appropriate choice. The latter exhibits a mixed valent phase in which valence fluctuations are dominant. However, they do not lead to any phase transition. To capture stronger effects of valence fluctuations, the Anderson Hamiltonian has been extended by including a Hubbard repulsion between localized $f$ and itinerant $c$ electrons in the Anderson model~\cite{Freericks,Hewson1}. In the literature, this term is  called the Falicov-Kimball term or the $U_{fc}$ term. We will use these terms interchangeably.

Several theoretical studies of the extended single-impurity Anderson model (e-SIAM) have been carried out. 
  In Ref.~\onlinecite{Hewson1}, the authors have carried out a numerical renormalization group (NRG) study of the e-SIAM, and they found that the $U_{fc}$ term does not lead to  significant effect on spectral and thermodynamic 
properties.
They could fit their results to the Anderson impurity model with renormalized parameters. In Ref.~\cite{Hewson2}, the authors have used  renormalized perturbation theory (RPT) to study the effect of $U_{fc}$ term and they found 
that there is no change in the low energy fixed point of the Anderson model due to this term and all it does is the renormalization of the parameters of Anderson impurity model. However, in Ref.~\onlinecite{Irkhin1}, a scaling analysis of SIAM with FK interaction showed that hybridization and hence the Kondo scale gets heavily renormalized. Later 
on, based on another NRG study of the asymmetric e-SIAM\cite{Irkhin2}, it was found that the FK interaction affects thermodynamic properties like specific heat. The authors have proposed that due to the FK interaction, there are excitonic excitations which lead to this renormalization. Yet another detailed NRG study has been reported in Ref.\onlinecite{Ryu}, where the effect of valence fluctuations on the excitation spectra of the model has been calculated. Since they have taken local Hubbard repulsion to be infinite, they find that spin fluctuations dominate the physics. However, they have proposed that for finite values of local Hubbard interaction, there will be contribution from charge Kondo effect as well. Thus, we find that the role of the $U_{fc}$ term and the ensuing valence fluctuations in the extended SIAM in still under debate. Hence, we have addressed this question by using a complementary set of methods which include perturbative renormalization methods as given in Refs.~\onlinecite{Anderson,Haldane,Jefferson} and the Schrieffer-Wolff transformation~\cite{SW, Hewson}.

There are additional reasons to study the e-SIAM. One important reason is that within dynamical mean field theory (DMFT), lattice models like the periodic Anderson model and its extended versions are mapped to the e-SIAM, so it becomes very important to understand the latter. Yet another reason to study the e-SIAM comes from a recent study of Ref.\onlinecite{Miyake2}.
The authors in the latter show that a charge Kondo effect can arise due to pair hopping mechanism. Since charge (valence) fluctuations play a significant role in quantum transport, the effect of $U_{fc}$ interaction has been studied in this context~\cite{Natan, Borda1, Borda2, Kuramoto}.
 Quantum criticality has also been found in impurity Anderson models with particular forms of the density of states~\cite{Pixley}. This gives an added motivation for this work. 
 
   In this paper, we have employed perturbative renormalization methods of Refs.~\onlinecite{Haldane, Jefferson} to study the scaling behaviour of e-SIAM, with a focus on the differences introduced by the $U_{fc}$ interaction term. Scaling
  trajectories of any Hamiltonian are governed by the scaling invariants of that model and hence to explore the effect of valence fluctuations in e-SIAM, we have calculated its scaling invariants. We find that they differ from those of SIAM. Since the strong coupling regime of the SIAM  is governed by Kondo physics, 
we have explored the renormalization of the Kondo scale
 due to valence fluctuations and one of the very important findings of this work is that Kondo scale of e-SIAM gets enhanced due to valence fluctuations. 
 It is known for the case of SIAM, that the hybridization does not get renormalized at second order level~\cite{Haldane,Hewson} and that is what we found for e-SIAM as well. Nevertheless, we wanted to explore the renormalization effects of $U_{fc}$ on hybridization so using Jefferson's method\cite{Jefferson} we did a third order scaling analysis of e-SIAM and calculated a scaling equation for the hybridization as well. Our perturbative renormalization calculations show that the
$U_{fc}$ interaction does have strong renormalization effects on the model parameters of e-SIAM, and hence the Kondo scale also gets renormalized. 

To explore these renormalization effects at an effective Hamiltonian level, we employed the Schrieffer-Wolff transformation and found that the strong coupling physics of e-SIAM is not governed by Kondo model rather it is the \textit{spin-charge Kondo model} which has an interplay of spin and charge Kondo effects. We also found that if one uses only projection operator method, what one gets is the standard Kondo model with renormalized Kondo coupling which does not capture the full effect of $U_{fc}$ interaction because projection operator method projects the Hamiltonian to the singly occupied subspace.  

The paper is  organized as follows: the model is introduced in the next section. Then, we have carried out a Schrieffer-Wolff transformation of the model (section 3) and we find that the effective Hamiltonian in the strong coupling regime of e-SIAM is an anisotropic spin-charge Kondo model in which both spin and charge Kondo interactions are present. Then, we have carried out a detailed perturbative renormalization study of the model
(sections 4 and 5) and calculated the scaling equations and the scaling invariants of the model. Analytical and numerical solutions of the scaling equations have been presented. We have done benchmarking of our results with NRG calculations and have found excellent agreement. Finally, we have summarized our results in section  6 and have concluded in the final section.

\section{Hamiltonian}
As the name suggests, the extended single impurity Anderson impurity (e-SIAM) model incoprporates an extension to the usual(standard) Anderson model 
in the form of a $U_{fc}$ term specifically added to capture the effect of enhanced valence fluctuations. In second quantized notation,
the Hamiltonian is written below:
\begin{align}
H=&\sum_{k\sigma}\epsilon_{k}c^{\dag}_{k\sigma}
c^{\phantom{\dag}}_{k\sigma}+
\sum_{\sigma}\epsilon_{d}d^{\dag}_{\sigma}d^{\phantom{\dag}}_{\sigma}+\sum_{k\sigma}V_{k}(c^{\dag}_{k\sigma}d^{\phantom{\dag}}_{\sigma}+d^{\dag}_{\sigma}c^{\phantom{\dag}}_{k\sigma})
\nonumber \\+
&Un_{d\uparrow}n_{d\downarrow}+\sum_{k\sigma\sigma'}
U_{fc}n_{k\sigma}n_{d\sigma'}\,.
\label{esiam-model}
\end{align}
The model captures the dynamics of a local impurity hybridising with a sea of free fermions which have dispersion but no interactions. These itinerant electrons are referred to as  $c$ electrons, and the first term corresponds to them. The impurity, which is a localized $d$ electron has no dispersion, but there is an 
on-site Hubbard repulsive interaction between $d$ electrons.The local impurity is represented by second and fourth terms. The hybridization between itinerant and localized electrons is written as the third term. The last term is the $U_{fc}$ term which captures the Hubbard repulsion between itinerant and localized electrons of the host and the impurity respectively. The standard SIAM ($U_{fc}=0$) has three main regimes called Kondo regime, the mixed valent regime and local moment regime which is connected by smooth crossovers. In the next section, we will carry out a unitary transformation which will yield an effective Hamiltonian and the difference between the cases of $U_{fc}=0$ and $U_{fc} > 0$ will become apparent.

\section{Effective Hamiltonian through a Schrieffer-Wolf transformation}
The Schrieffer-Wolff transformation(SWT) is a method which gives the low energy effective Hamiltonian of a given quantum many-body Hamiltonian by projecting out the high energy excitations. In case of SIAM, this transformation maps the model  to Kondo model which lies at strong coupling fixed point of SIAM. To understand the physics of the e-SIAM, we have again employed SWT and calculated the corresponding effective Hamiltonian. There are at least two different ways of doing SWT: 1)One can use unitary transformation method as used in Ref.~\onlinecite{SW} or 2) One can use projection operator method as in Ref.~\onlinecite{Hewson}. We have used both of these methods, and we will be pointing out an advantage of the former with respect to the latter. 

The generator of SW transformation for the e-SIAM is given by
\begin{equation}
S = \sum_{k\sigma}(A_{k} + B_{k}n_{d\bar{\sigma}})V_{k}(c^{\dag}_{k\sigma}d_{\sigma}-d^{\dag}_{\sigma}c_{k\sigma}) 
\end{equation}
where $A_{k}$ and $B_{k}$ are given below:
\begin{align}
A_{k}& =\frac{1}{\epsilon_{k}-\epsilon_{d}}\\
B_{k}& = \frac{1}{\epsilon_{k}-\epsilon_{d}+U_{fc}-U} - \frac{1}{\epsilon_{k}-\epsilon_{d}}
\end{align}
To carry out the transformation, we evaluated the commutator,$[S,H_{v}]$ (where $H_v$ is the hybridization term in the Hamliltonian, equation~\ref{esiam-model}), as given below:
\begin{align}
&[S,H_{v}] =\nonumber \\
&\sum_{kk'\sigma}A_{k}V_{k}V_{k'}(c^{\dag}_{k\sigma}c_{k'\sigma})-\sum_{k\sigma}A_{k}V_{k}^{2}(d^{\dag}_{\sigma}d_{\sigma})- \nonumber \\
&\sum_{k\sigma}B_{k}V_{k}^{2}(n_{d\bar{\sigma}}d^{\dag}_{\sigma}d_{\sigma})
-\sum_{kk'\sigma}B_{k}V_{k}V_{k'}(c^{\dag}_{k'\bar{\sigma}}d_{\bar{\sigma}}c^{\dag}_{k\sigma}d_{\sigma}) +
\nonumber \\
&\sum_{kk'\sigma}B_{k}V_{k}V_{k'}(d^{\dag}_{\bar{\sigma}}c_{k'\bar{\sigma}}c^{\dag}_{k\sigma}d_{\sigma}) +\sum_{kk'\sigma}B_{k}V_{k}V_{k'}(c^{\dag}_{k\sigma}c_{k'\sigma}n_{d\bar{\sigma}})\nonumber \\
&+h.c.
\end{align}
We need to switch to Nambu spinor notation to write the Kondo exchange term in terms of spin operators. There are other terms also present in $[S,H_{v}]$ including the $H_{ch}$ which is the longitudinal part of the charge Kondo interation.
 Combining  Kondo exchange terms with $H_{ch}$ and $U_{fc}$ term from $H_{0}$ we get the following effective Hamiltonian.
\begin{align}
H_{eff}=&\sum_{k}\epsilon_{k}c_{k\sigma}^{\dag}c_{k\sigma}+U_{fc}\sum_{k}n_{k\sigma}n_{d\sigma'}+H_{ex}+
H_{ch} \\
&H_{ex}=\sum_{kk'\sigma}J_{kk'}\left(\Psi^{\dag}_{k}S\Psi_{k'}\right) \left(\Psi^{\dag}_{d}S\Psi_{d}\right) \\
&H_{ch}=\frac{1}{2}\sum_{kk'\sigma}J_{kk'}\left(c^{\dag}_{k\bar{\sigma}}d_{\bar{\sigma}}
c^{\dag}_{k'\sigma}d_{\sigma}\right)+h.c.
\end{align}
where  $J_{kk'}$ and $W_{kk'}$ are given by:
\begin{align}
J_{kk'}&=V_{k}V_{k'}\Big(\frac{1}{\epsilon_{k}-\epsilon_{d}+U_{fc}-U}
+\frac{1}{\epsilon_{k'}-\epsilon_{d}+U_{fc}-U}\nonumber \\
&-\frac{1}{\epsilon_{k}-\epsilon_{d}}-
\frac{1}{\epsilon_{k'}-\epsilon_{d}}\Big)\\
W_{kk'}&=V_{k}V_{k'}\left(\frac{1}{\epsilon_{k}-\epsilon_{d}}+\frac{1}{\epsilon_{k'}-\epsilon_{d}}\right)
\end{align}
Choosing $k=k'$, the Kondo exchange becomes
\begin{align}
J_{k}=2V^{2}_{k}\left(\frac{1}{\epsilon_{k}-\epsilon_{d}+U_{fc}-U}-\frac{1}{\epsilon_{k}-\epsilon_{d}}\right) 
\end{align}
In the isospin representation, the $U_{fc}$ term is the longitudinal  component of the charge Kondo interaction which is $I_{c}^{z}I_{d}^{z}=(n_{k\sigma}-1)(n_{d\sigma}-1)$ where $I_{c}$ and $I_{d}$ are isospin operators of conduction electrons and impurity respectively. This longitudinal charge Kondo interaction term($U_{fc}$) is present in the effective Hamiltonian (in $H_{0}$) which when combined with $H_{ch}$ gives the full charge Kondo interaction, which, in contrast to the isotropic spin Kondo interaction, is anisotropic in nature. 
Thus, the full effective Hamiltonian that we have obtained from SWT of e-SIAM is the spin-charge Kondo model(SCKM). $H_{ch}$ is usually ignored by arguing that spin Kondo model lives in $n_{d}=1$ subspace of Anderson Hamiltonian. However,  as noted by Salomaa~\cite{Salomaa}, in a system with valence fluctuations both Kondo interactions are significant. From symmetry point of view, charge Kondo interaction has $su(2)_{c}$ symmetry which commutes with symmetry of spin Kondo interaction~\cite{Zitko3}. Charge Kondo interaction interacts with isopsin(pairing) part of the conduction bath and gives rise to Kondo effect~\cite{Taraphder}. So we have obtained the spin-charge Kondo model in which spin and charge Kondo effects co-exist as was been already found in NRG calculations~\cite{Miyake2}. Our results show that  spin-charge Kondo model can arise in a system with repulsive interactions alone and there is no need for phononic mechanisms to have attractive interaction.

\subsection{Effective Hamiltonian through projection operator method}
A minor point that we would like to emphasise is that the elimination of 
charge fluctuations can be done in multiple ways, and there are subtle differences between the methods. For example, the projection operator method does yield the Kondo model, when applied to the conventional SIAM. However, since the projection to the $n_d=1$ subspace is built into the method, the charge Kondo terms automatically vanish. This is
in contrast to the SWT, which is a unitary transformation and until a projection to the singly occupied subspace is carried out, all the quartic
operators remain. This implies that the projection operator method, yields 
only a renormalized Kondo model (ignoring potential scattering), 
\begin{equation}
H=\sum_{k\sigma}\sum_{k\sigma'}(\epsilon_{k}c_{k\sigma}^{\dag}c_{k\sigma}+
J_{kk'}S.c_{k\sigma}^{\dag}(\sigma)_{\sigma\sigma'}c_{k'\sigma'}
\end{equation}
where the coupling constant is given by
\begin{equation}
J_{kk'}=-V_{k}V_{k'}\left(\frac{1}{U+\epsilon_{d}-\epsilon_{k'}-2U_{fc}}+
\frac{1}{\epsilon_{k}-\epsilon_{d}}\right)
\end{equation}
and not the full spin-charge Kondo model given by the SWT.
In the next two sections, namely sections 4 and 5, we will carry out perturbative scaling studies of the e-SIAM for finite $U$ and
infinite $U$ limits respectively for the particle-hole symmetric and asymmetric cases.
 
\section{Perturbative scaling of the E-SIAM: Finite $U$}
 In this section, we will apply perturbative renormalization methods \cite{Anderson,Haldane,Jefferson,Hewson}to the e-SIAM. Our focus will be on the changes in the scaling equations and the corresponding scaling invariants due to the $U_{fc}$ interaction. 
 We begin with a calculation of the renormalization of the impurity energy levels. This will allow us to extract the scaling equations for the orbital energy, $\epsilon_d$ and the Hubbard $U$. Following the same procedure of poor man scaling for e-SIAM as done for Anderson impurity model~\cite{Haldane,Hewson}, we obtain the renormalized impurity energy levels:
\begin{align}
&E'_{0}=E_{0}-\frac{2\bigtriangleup}{\pi}\frac{\mid \delta D \mid}{D+\epsilon_{d}}\\
&E'_{1}=E_{1}-\frac{\bigtriangleup\mid \delta D\mid}{\pi}\Big(\frac{1}{D+U_{fc}-\epsilon_{d}}+\frac{1}{D+2U_{fc}+\epsilon_{d}+U}\Big)\\
&E'_{2}=E_{2}-\frac{2\bigtriangleup\mid \delta D\mid}{\pi}\Big( \frac{1}
{D+2U_{fc}-\epsilon_{d}-U}\Big)
\end{align}
where $E_{0}$,$E_{1}$,$E_{2}$ are the energies of empty, singly occupied and doubly occupied impurity electron states. Given these renormalized energies,
the scaling equations for the interaction strength and the orbital energy may be obtained through $\epsilon_d=E^\prime_1 - E^\prime_0$
and $U=E^\prime_2 - 2E^\prime_1 + E^\prime_0$ as:
\begin{align}
\frac{dU}{dD} =&-\frac{2\Delta}{\pi}\Big(\frac{1}{D+U_{fc}-\epsilon_d}
+\frac{1}{D+2U_{fc}+\epsilon_d+U} -\nonumber \\
& \frac{1}{D+2U_{fc} -\epsilon_d - U}
-\frac{1}{D+\epsilon_d}\Big) \\
\frac{d\epsilon_{d}}{dD} =&-\frac{\Delta}{\pi}\Big(\frac{2}{D+\epsilon_d}-\frac{1}{D+U_{fc}-\epsilon_d} +\nonumber \\
& \frac{1}{D+2U_{fc}+\epsilon_d+U} \Big)
\label{eq:hal_ed}
\end{align}
These equations may be solved easily using the Euler's discretization,
and the results are presented below. 
\subsection{Particle-hole symmetric case}
\begin{figure}
\centering
\includegraphics[clip,scale=0.5]{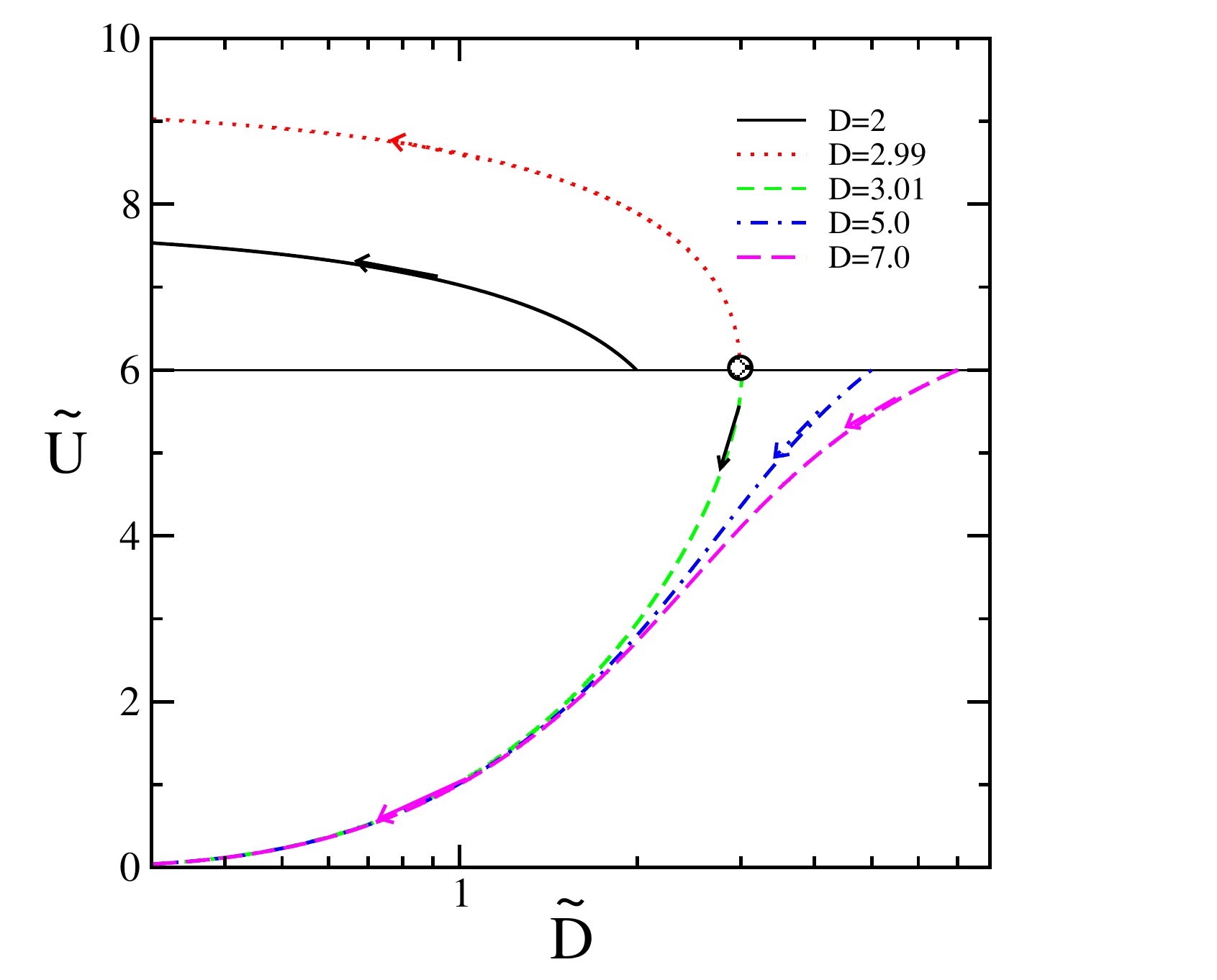}
\caption{Scaling flow of $U$ for the particle-hole symmetric case with $U=6, \epsilon_d=-U/2$, $\Delta=1$ for $U_{fc}=0$.}
\label{eta0-U-scaling}
\end{figure}

For the particle-hole (ph)
symmetric case, $\epsilon_d=-U/2$ is maintained even during the
flow, hence we can focus only on the interaction strength. With 
$U_{fc}=0$, the flow should be the same as that of the conventional Anderson
model. So, we study the solutions of the scaling equations for $U_{fc}=0$ first in order to understand the limitations and the merits of the perturbative scaling. Figure~\ref{eta0-U-scaling} shows the flow of $U$ with decreasing 
bandwidth for various initial $D$ values. We see that if the orbital
energy is within the initial band, the interaction strength flows to 
lower values and eventually vanishes, implying a flow to a non-interacting system and hence the strong coupling fixed point. While if the initial bandwidth, $D$ is smaller than $|\epsilon_d|$, the interaction strength
flows to higher values, and this may be interpreted as a flow to the local moment fixed point. Hence the separatrix is $D=-\epsilon_d$. It is well known
(e.g from NRG calculations) that a separatrix is absent in the SIAM, and
for all initial values in the ph-symmetric case, the system flows to
the strong coupling fixed point. The flow to the LM fixed point seen in figure~\ref{eta0-U-scaling} is an artefact of the perturbative renormalization
employed here. Next, we investigate the effect of $U_{fc}$ on the scaling
trajectories. 
For $U_{fc}=0$, we have seen from figure~\ref{eta0-U-scaling} that for $D<-\epsilon_d$, the flow
is always towards the LM fixed point, since the $U$ increases monotonically with decreasing bandwidth.
The Falicov-Kimball interactions changes the flow as shown below.
The top panel of figure~\ref{eta0-U-Ufc} shows that for the same parameter regime 
( $D=2.5$, such that $D < -\epsilon_d$), a new separatrix is introduced at a finite value of $U_{fc}$,
which separates the upward renormalization from the downward flow. However, a difference with the
$U_{fc}=0$ case is that the $D\rightarrow 0 $ value of $U$ is finite instead of vanishing as in figure~\ref{eta0-U-scaling}.
Nevertheless, as $U_{fc}$ increases, the system always flows towards lower interaction strengths, implying
an increase in valence fluctuations. This observation is reiterated in the bottom panel of 
figure~\ref{eta0-U-Ufc} ($D=3.1$ such that $D>-\epsilon_d$),
where increasing $U_{fc}$ leads to uniformly downward renormalization and progressively smaller 
values of the interaction strength as $D\rightarrow 0$. 
\begin{figure}
\includegraphics[clip,scale=0.5]{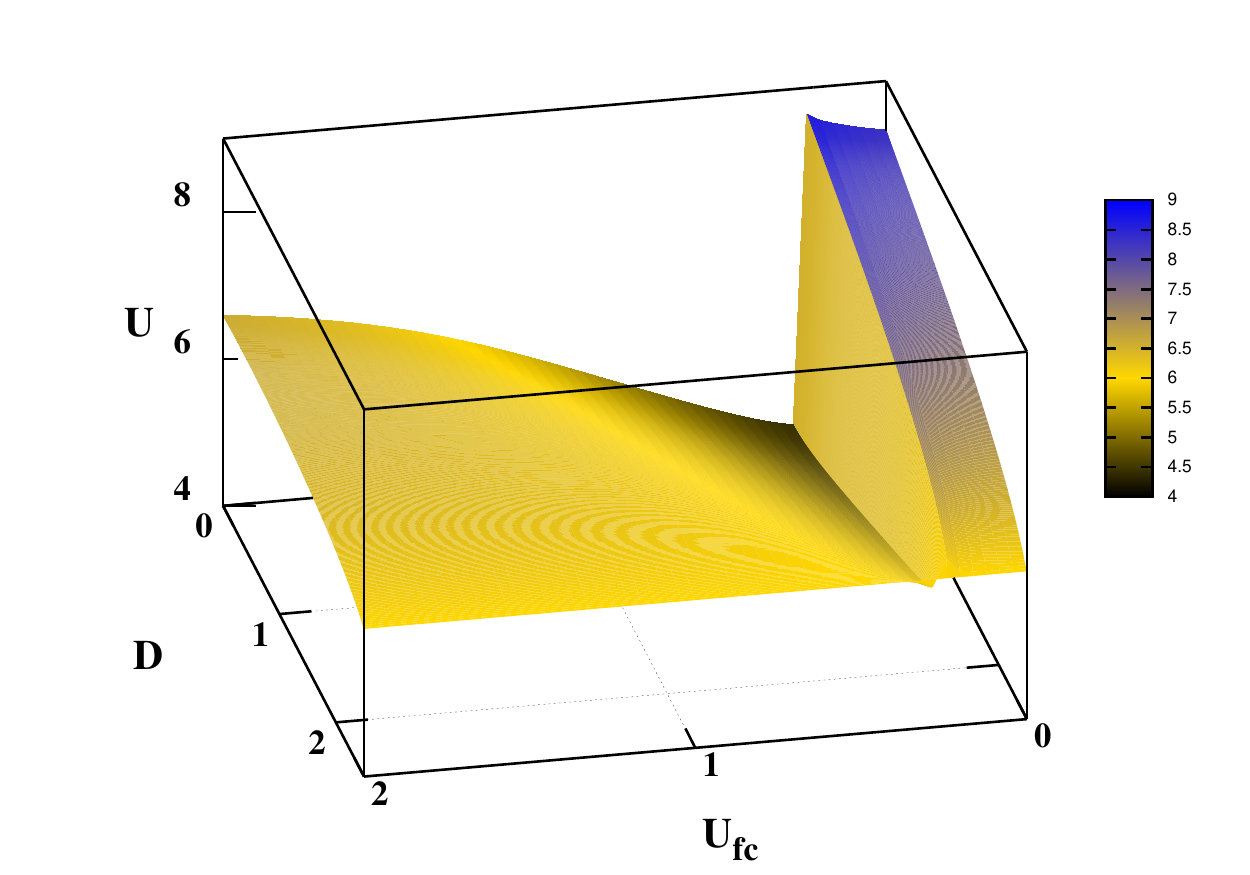}
\includegraphics[clip,scale=0.5]{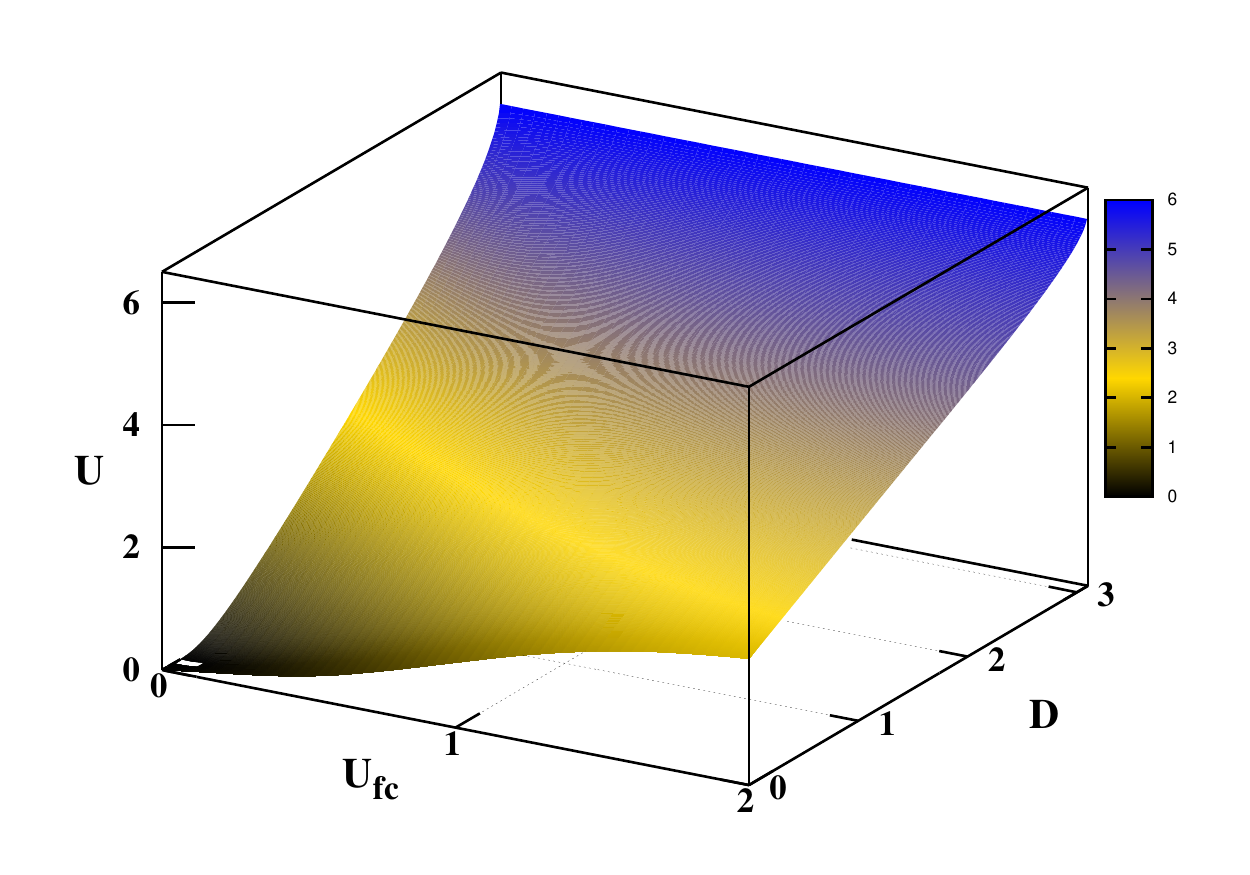}
\caption{Scaling flow of $U$ with decreasing bandwidth, for various $U_{fc}$ values in the ph-symmetric case. The initial bandwidth 
in the top panel is $D=2.5$, such that $D < -\epsilon_d$ and in the bottom panel, $D=3.1$, such that $D>-\epsilon_d$.}
\label{eta0-U-Ufc}
\end{figure}
\subsection{Particle-hole asymmetric case}
\begin{figure}
\centering
\includegraphics[clip,scale=0.6]{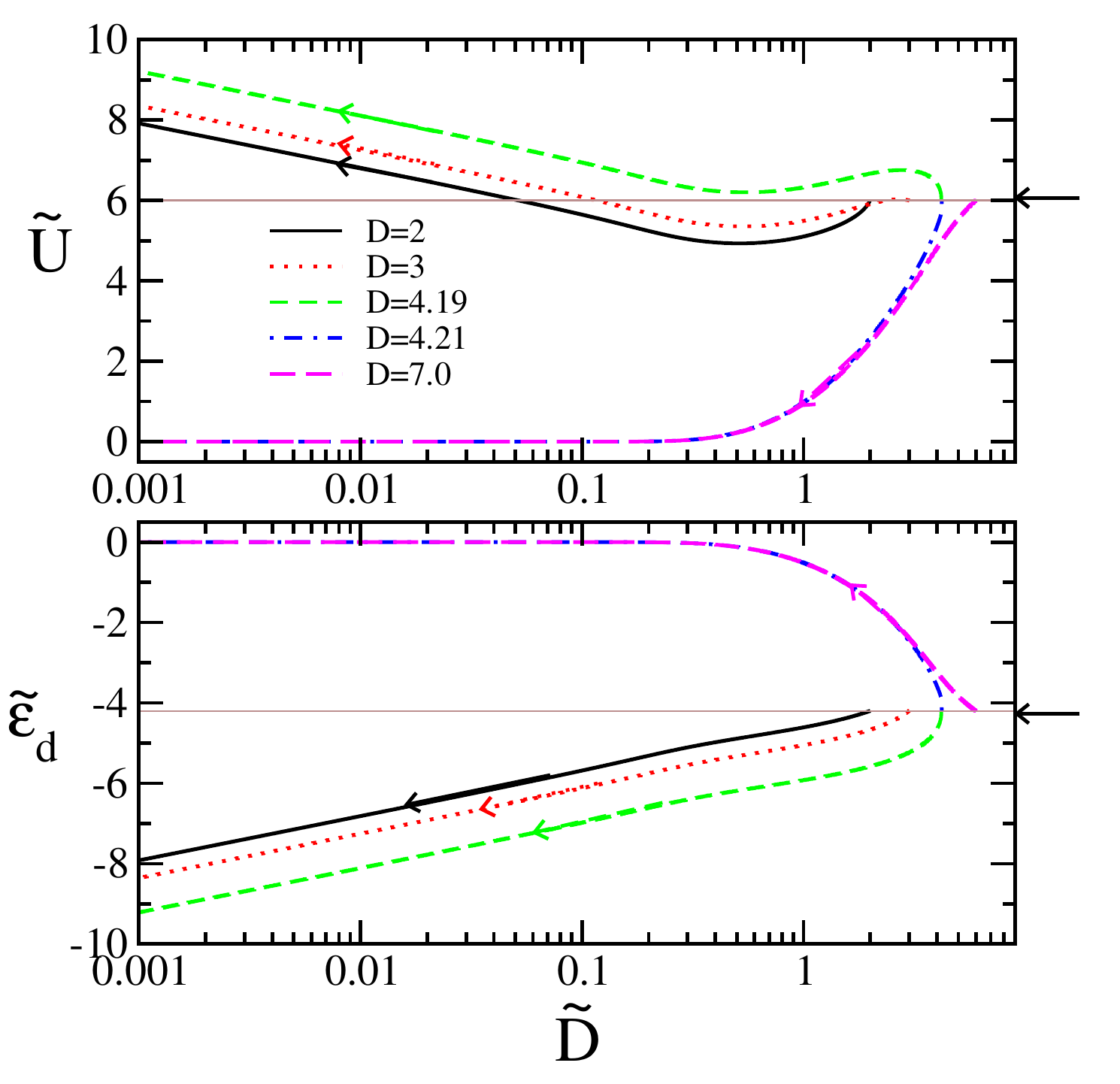}
\caption{Scaling flow of U(top panel) and $\epsilon_{d}$(bottom panel) for asymmetric case with $U_{fc}=0$. Other parameters used for this figure are: $U=6.0$, $\epsilon_{d}=-4.2$ $\eta=-0.4$}
\label{phasym_U_ed_flow_Ufc0}
\end{figure}
In the p-h symmetric case, the $d$-occupancy is 1, and the asymmetry, defined as $\eta=1+2\epsilon_d/U$
vanishes. In the asymmetric case, $\epsilon_d\neq -U/2$, hence $\eta\neq 0$ and the occupancy $n_d$ deviates from unity,
 becoming either electron doped ($n_d > 1$)
or hole doped ($n_d<1$). Before we investigate the effects of $U_{fc}$ for $\eta\neq 0$, the behaviour of the scaling equations
for $U_{fc}=0$ should be understood. We show the scaling flow of $U$ and $\epsilon_d$ for decreasing bandwidth in the top
and bottom panels respectively of figure~\ref{phasym_U_ed_flow_Ufc0}. In contrast to the symmetric case, the flows
here are seen to be quite non-monotonic and interesting. Although $D=-\epsilon_d$ is still a separatrix, the $D<<|\epsilon_d|$
flows show initial downward renormalization, but as $D\rightarrow 0$, the interaction strength grows and saturates at a finite
value higher than the initial value. For all $D >|\epsilon_d|$ however, the $\tilde{U}$ eventually vanishes. So, qualitatively,
the infrared flows are exactly the same for the symmetric and the asymmetric case. The lower panel of figure~\ref{phasym_U_ed_flow_Ufc0} showing $\tilde{\epsilon}_d$ mirrors the flows seen for the symmetric case. For
an initial impurity energy within the band, the flow is towards $\tilde{\epsilon}_d\rightarrow 0$, while for initial $\epsilon_d$ below
the band, the renormalization is towards the LM fixed point. Next we turn on $U_{fc}$.
\begin{figure}
\includegraphics[clip,scale=0.5]{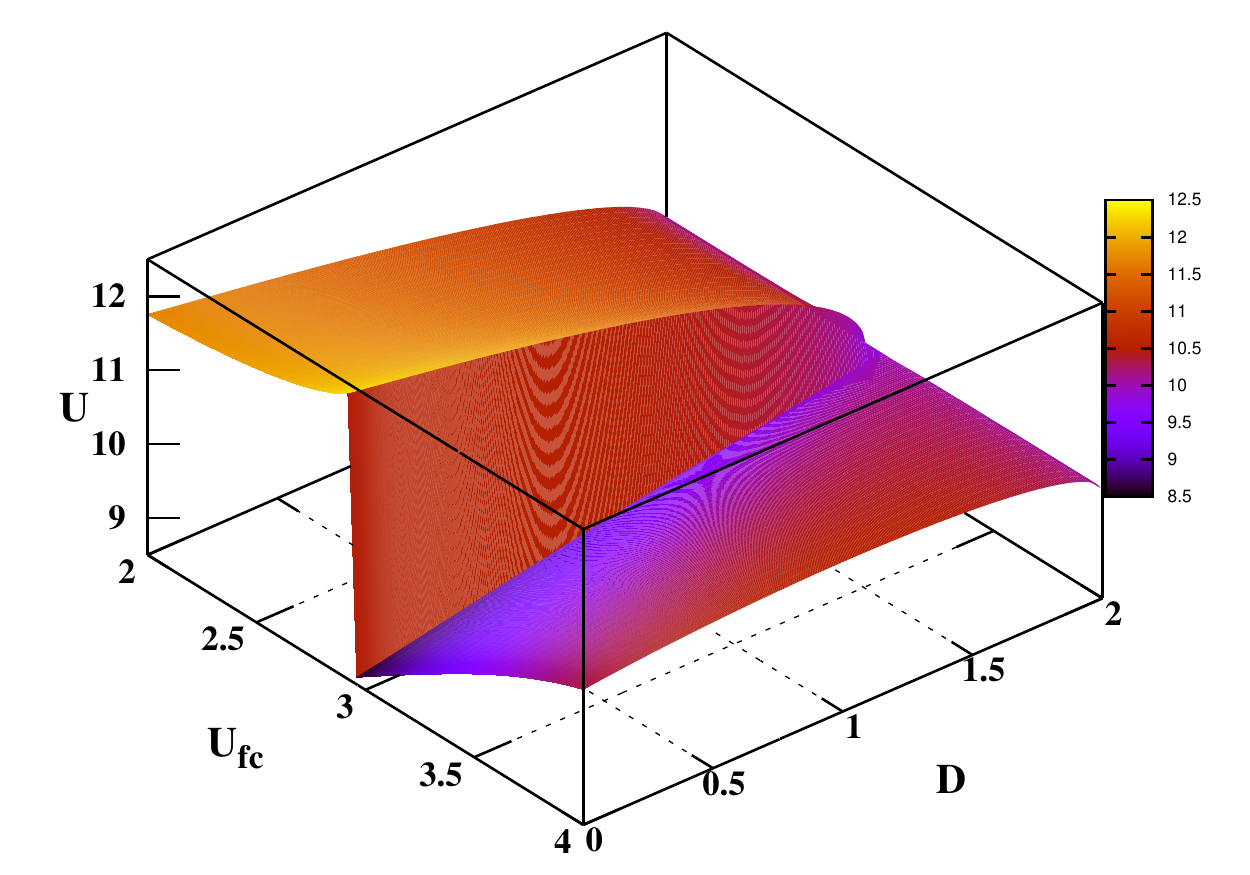}
\includegraphics[clip,scale=0.5]{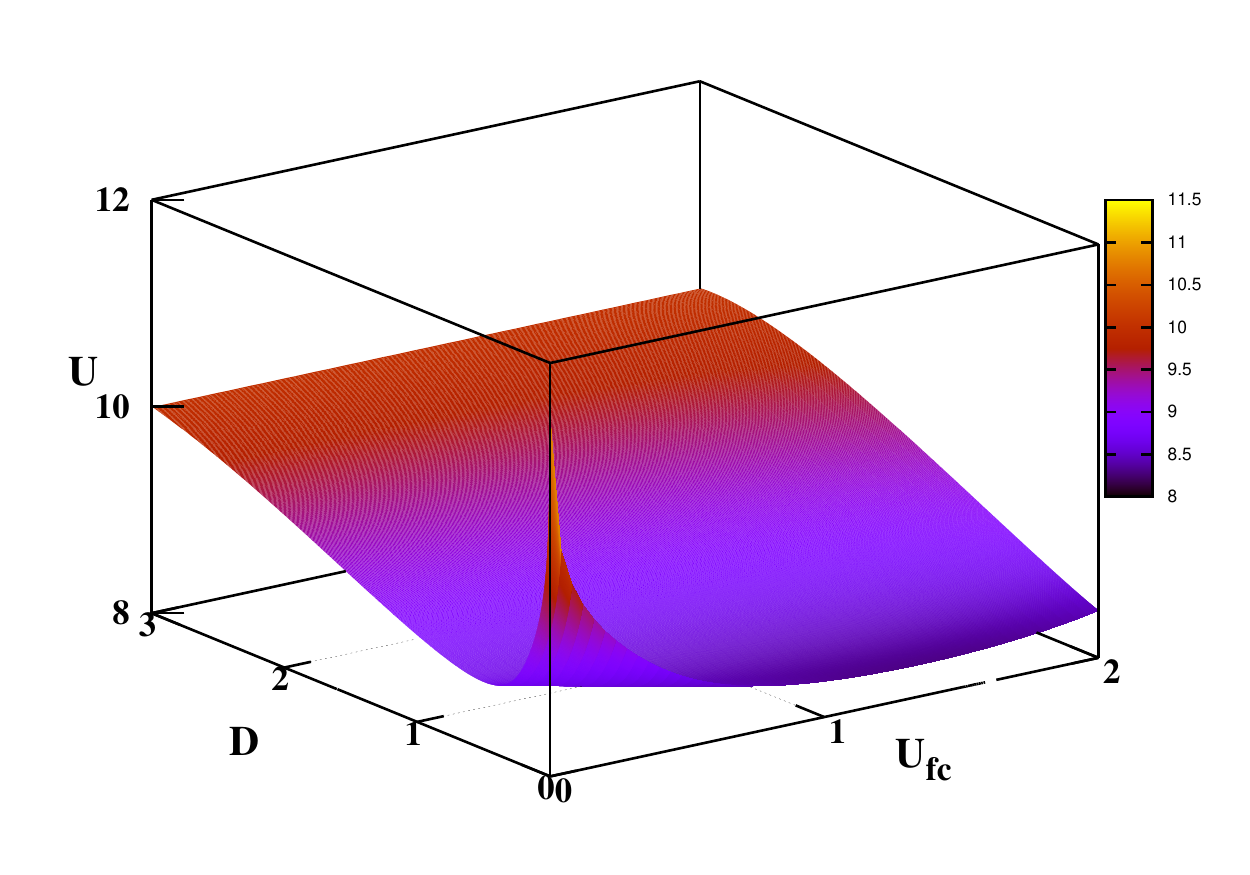}
\caption{Scaling flow of $U$ with decreasing bandwidth for various $U_{fc}$ values in the ph-asymmetric case with initial asymmetry
of $0.58$. The initial bandwidth in the top panel is $D=2.0$ such that $D < -\epsilon_d$ and in the bottom panel, $D=3.0$, such that $D>-\epsilon_d$. The initial interaction strength is $U=10.0$}
\label{eta0.58-U-Ufc}
\end{figure}
The effect of $U_{fc}$ is expected to be far more significant in the p-h asymmetric case, since valence
fluctuations are much more favourable when $n_d\neq 1$.  Figure~\ref{eta0.58-U-Ufc} shows the flow of $U$ with 
decreasing bandwidth for $U=10.0$, and $\eta=0.58$, which implies $\epsilon_d=-2.1$. The top panel shows results
for $D< -\epsilon_d$, such that the $d$-level lies below the conduction band. In the absence of $U_{fc}$, the system
flows to the LM fixed point, but with increasing $U_{fc}$, a separatrix is introduced at $U_{fc}\sim 2.9$ in parallel
to the p-h symmetric case. Since the $U$ flows downward beyond this separatrix, we can interpret this as $U_{fc}$
driven increase in valence fluctuations. For, $D>-\epsilon_d$, such that the $d$-level lies within the conduction band, the
flow in the absence of $U_{fc}$ (as seen in the bottom panel of figure~\ref{eta0.58-U-Ufc}) is non-monotonic. The $U$ decreases initially,
bu as the bandwidth reduces, the interaction strength goes through a shallow dip and increases steeply as $D\rightarrow 0$.
Again this behaviour changes qualitatively with increasing $U_{fc}$. Although there is no separatrix, the $U$ decreases 
monotonically for larger $U_{fc}$ again leading to the interpretation that valence fluctuations are enhanced.

In the next section, we will consider the $U\rightarrow \infty$ limit so that we will study the effect of valence fluctuations
between the empty and singly occupied states only.
\section{Perturbative scaling of the E-SIAM: Infinite $U$ limit}
We consider the  $U\rightarrow \infty$ limit due to which doubly occupied states get decoupled so 
the renormalization of Hubbard repulsion is not a consideration in the following. In this section, we will
first get the scaling equation for $\epsilon_d$, and subsequently investigate the renormalization of hybridization.
\subsection{Scaling flow of $\epsilon_d$}
\label{sec:reed}
\begin{figure}
\centering
\includegraphics[clip=,scale=0.6]{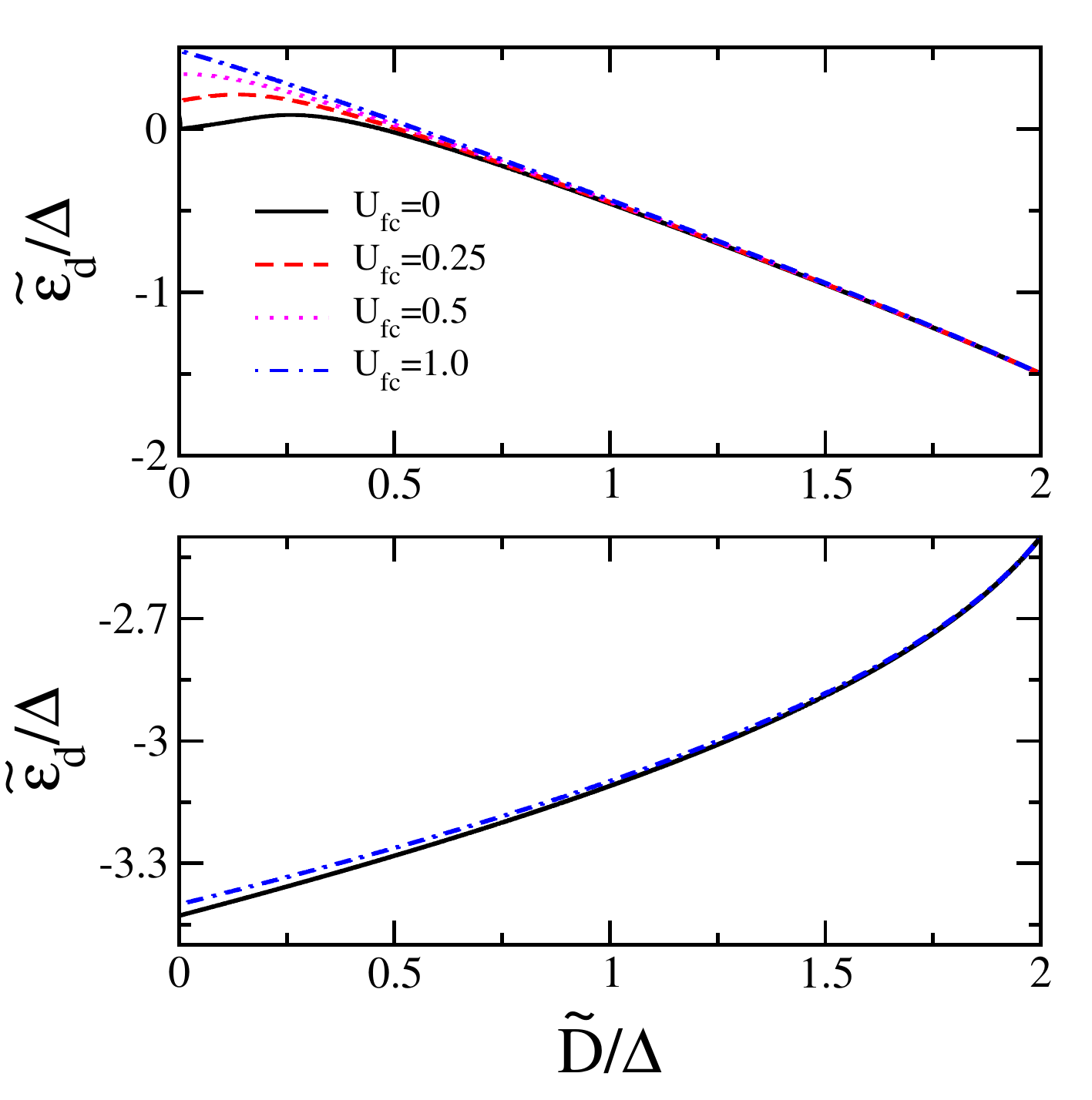}
\caption{Scaling flow of $\epsilon_{d}$ for mixed valent regime($D> -\epsilon_d$, top panel) and local moment regime
 ($D> -\epsilon_d$, bottom panel).Initial value of $\epsilon_{d}= -1.5$ for top panel and $\epsilon_{d}=-2.5$ for bottom panel. Other parameters are: $D=2$, $\Delta=1$}
\label{flow_ed_u_infty}
\end{figure}

Following Jefferson~\cite{Jefferson} we will calculate the scaling equation for $\epsilon_{d}$ by calculating the effective Hamiltonian till second  order and comparing with the bare Hamiltonian given in equation~\ref{esiam-model}.

\begin{align}
H(\tilde{D})&=\sum_{k\sigma}\epsilon_{k}c^{\dag}_{k\sigma}c_{k\sigma}+\sum_{k\sigma\sigma'}U_{fc}
n_{k\sigma}n_{d\sigma'}+\nonumber \\
&\sum_{k}V_{k}^{0}(c^{\dag}_{k\sigma}d_{\sigma}+d_{\sigma}^{\dag}c_{k\sigma})+\nonumber\\
&\sum_{\sigma}(\epsilon_{d}^{0}+\frac{\rho_{0}V^{2}\delta D}{-D-U_{fc}+\epsilon_{d}}-
\frac{2\rho_{0}V^{2}\delta D}{-D-\epsilon_{d}})n_{d\sigma}\,,
\end{align}
where $V^{0}_{k}$ and $\epsilon_{d}^{0}$ are bare values of hybridization and impurity orbital energy and $k$ is restricted to model space. It is easy to see that the impurity energy has got renormalized. After comparing with the bare Hamiltonian(equation~\ref{esiam-model}) we can write
 down the effect due to renormalization.
\begin{align}
\delta \epsilon_{d}=\rho_{0}V^{2}\delta D(\frac{2}{D+\epsilon_{d}}-\frac{1}{D+U_{fc}-\epsilon_{d}})
\end{align}
This equation is identical to the one obtained before, equation~\ref{eq:hal_ed} in the limit of $U\rightarrow \infty$.
We have solved this equation numerically and the results are presented  in figure~\ref{flow_ed_u_infty}.
As is shown in figure~\ref{flow_ed_u_infty}, the effect of $U_{fc}$ is stronger in the mixed valent regime ($D > -\epsilon_d$, top panel) while for the local moment regime ($D < -\epsilon_d$, bottom panel), where the impurity energy level lies deeper below the band, the effect of $U_{fc}$ is insignificant. A few analytical forms may be obtained in a limiting case, namely $D\gg |\epsilon_d|$.
The scaling equation for $\epsilon_{d}$ becomes in this case:
\begin{align}
\frac{d\epsilon_{d}}{dD}=-\frac{\Delta}{\pi}\left(\frac{2}{D}-\frac{1}{D+U_{fc}}\right)\,,
\end{align}
where $\Delta=\pi\rho_0 V^2$.
A scaling invariant can be obtained through this equation.
\begin{align}
\epsilon_{d}+\frac{\Delta}{\pi}\left[ln\frac{D}{D_{0}}+ln\left(\frac{D}{D_{0}}\frac{D_{0}+U_{fc}}
{D+U_{fc}}\right)\right]=const
\end{align}
Using the fact that $U_{fc}$ is small as compared to bandwidth $D$ which is the largest enery scale of the model, we can further simplify this expression of scaling invariant.
\begin{align}
\epsilon_{d}^{*}=\epsilon_{d}+\frac{\Delta}{\pi}ln\frac{D}{D_{0}}-\frac{\Delta}{\pi}\frac{U_{fc}}{D}
\end{align}
The scaling invariant of e-SIAM has been written in this form to see its relation with the corresponding scaling invariant for SIAM where $U_{fc}=0$. The first two terms constitute the scaling invariant for SIAM and third term gives the contribution of $U_{fc}$ term.  From the scaling analysis of SIAM~\cite{Haldane,Jefferson,Hewson}, it is known that $\epsilon_{d}$ increases with scaling and the impurity energy level moves closer to the Fermi level and hence inreasing the valence fluctuations. From the above equation we can see that the effect of $U_{fc}$ term is to further enhance the increase in $\epsilon_{d}$ and hence $U_{fc}$ term enhances the valence fluctuations even further.

We have already calculated first  \textit{scaling invariant} for our model. At second order, there is no renormalization of hybridization so 
$\Delta$ is another scaling invariant. In our model, we have third scaling invariant as well which is $U_{fc}$ interaction itself. So the renormalization flow of e-SIAM is characterized by three scaling invariants. Later on, we will see the hybridization gets renormalized and we will find out the scaling behaviour of hybridization at the third order level.
\subsection{Renormalization of Hybridization}
\label{sec:rehyb}
The hybridization does not get renormalized at the second order level and hence is taken as a scaling invariant for the Anderson impurity model~\cite{Haldane}. In the e-SIAM also, we did not get any renormalization of hybridization at the second order of perturbative renormalization and hence hybridization is once again a scaling invariant of this model. However at third order, hybridization does get renormalized and for the conventional Anderson impurity model, Jefferson has calculated the corresponding scaling 
equations~\cite{Jefferson}. In this section we will calculate the scaling equations for hybridization 
in the e-SIAM. Here also we will continue to keep doubly occupied state decoupled and hence there will be no contributions of
 the processes to/from that state. The third order contributions to the effective Hamiltonian are given by:
\begin{align}
&H_{v}(\tilde{D})=\nonumber\\
&(1-P_{\delta D})H_{v}\sum_{\alpha}G_{\alpha}H_{v}G_{\alpha}H_{v}(1-P_{\delta D}^{\alpha}) \nonumber\\
&-(1-P_{\delta D})H_{v}\sum_{\alpha}G_{\alpha}(\sum_{\alpha'}G_{\alpha'}H_{v}(1-P_{\delta D}^{\alpha'})H_{v}(1-P_{\delta D}^{\alpha})
\end{align}
where $G_{\alpha}=\frac{P_{\delta D}}{E_{\alpha}-H_{0}}$ is the projected resolvent and $\alpha,\alpha'$ are the indices for the degenerate states. Since we have excluded the doubly occupied state,  the first term in the above equation will not contribute. So 
the second term is the only third order contribution to the effective Hamiltonian. To get the scaling equation, we need to calculate
this term for our model. We will see that the two terms of the hybridization  ($c^\dag_{k\sigma}d^{\phantom{\dag}}_\sigma$ and its Hermitian conjugate) get renormalized in different ways so we write them as follows and find the scaling equations separately for them.
\begin{align}
H_{v}=\sum_{k\sigma}V_{k1}c^{\dag}_{k\sigma}d_{\sigma}+V_{k2}d^{\dag}_{\sigma}c_{k\sigma}
\end{align}
 Using the fact there are no particles/holes in high energy states and summing over  the intermediate states, we get the renormalized hybridization expressions as follows:
\begin{align}
V_{2}-V_{0}&=\frac{-\rho_{0} V_{1}V_{2}^{2}\delta D}{(D+U_{fc}-\epsilon_{d})(D-\epsilon_{k})}\\
V_{1}-V_{0}&=\frac{-2\rho_{0} V_{2}V_{1}^{2}\delta D}{(D+\epsilon_{d})(D+\epsilon_{k})}
\end{align}
The scaling equations for the hybridization can then be written from the above equations:
\begin{align}
\frac{d V_{2}}{dD} &=\frac{\rho_{0}V_{1}V_{2}^{2}}{(D+U_{fc}-\epsilon_{d})(D-\epsilon_{k})}
\label{eq:v2}\\
\frac{d V_{1}}{dD}&=\frac{2\rho_{0} V_{2}V_{1}^{2}}{(D+\epsilon_{d})(D+\epsilon_{k})}
\label{eq:v1}
\end{align}
We will first present two analytically tractable limits for these equations, which will provide qualitative
insight and the third scaling invariant.

I. In the mixed valent regime, and close to the Fermi level, we can choose
$\epsilon_{k}=0, \epsilon_{d}=0$. In this regime also, we have two limits. First when $D>> U_{fc}$ in which case the scaling equation reduces to that of the SIAM as given in Ref.~\onlinecite{Jefferson}. In the second case, when $U_{fc}$ is comparable to the bandwidth, the
 scaling equations have different solutions and are given below. Also if we divide the two scaling equations for hybridizations and integrate,
 we find that $V_{1}$ and $V_{2}$ are related as $V_{1}=\frac{V_{2}^{2}}{V_{0}}$.  This relation implies that $V_2$ and $V_1$
 renormalize in exactly the same way in this limit ($\epsilon_d=0$ or $|\epsilon_d| \ll D$). It also needs to be noted that we have ignored
 the momentum dependence of the hybridization amplitudes which is physically reasonable in this regime because of the closeness 
 to Fermi level. So the scaling equation for $V_{2}$ becomes:
\begin{align}
\frac{dV_{2}}{dD}=\frac{V_{2}^{2}V_{1}\rho_{0}}{D(D+U_{fc})}
\end{align}
Solving for hybridization we obtain:
\begin{align}
\frac{1}{V_{2}^{3}}-\frac{1}{V_{0}^{3}}=\frac{-\rho}{4V_{0}U_{fc}}ln\left(
\frac{D}{D_{0}}\frac{D_{0}+U_{fc}}{D+U_{fc}}\right)
\end{align}
Once again using the fact that bandwidth is the largest energy scale of the model,we arrive at the simplified expression for hybridization.
\begin{align}
\frac{1}{V_{2}^{3}}=\frac{-\rho}{4 V_{0}}\left(\frac{1}{D_{0}}-\frac{1}{D}\right)
+\frac{U_{fc}\rho}{4V_{0}}\left(\frac{1}{D_{0}}-\frac{1}{D}\right)^{2}
\label{eq:mv_v2}
\end{align}
We notice that $U_{fc}$ enters this expression for hybridization at second order. For the case of vanishing $U_{fc}$, we arrive at following scaling invariant for Anderson impurity model.
\begin{align}
\frac{1}{V_{2}^{3}}-\frac{\rho_{0}}{4V_{0}D}=\frac{1}{V_{0}^{3}}-\frac{\rho_{0}}{4V_{0}D_{0}}
\end{align}
Solving for $V_{2}$ we arrive at the following equation which was obtained by Jefferson~\cite{Jefferson}.
\begin{align}
V_{2}=V_{0}\left(1+\frac{\rho V_{0}^{2}}{4D_{0}}\left(\frac{D_{0}}{D}-1\right)\right)^{-1/3}
\end{align}
As was shown by Jefferson for $U_{fc}=0$, the hybridization becomes weaker under the scaling flow. 
The first term on the RHS of equation~\ref{eq:mv_v2} shows that as $D\rightarrow 0$, $V_2$ also
decreases. The second term is the contribution of $U_{fc}$, which is seen to enhance the reduction
of $V_2$ even further.  Combined with the fact that $|\epsilon_d| \ll D$ implies a flow to the empty orbital
regime (see figure~\ref{flow_ed_u_infty}), we deduce that in presence of $U_{fc}$, valence fluctuations get enhanced.

II. In the LM regime, i.e $|\epsilon_d| \gg D$, equations~\ref{eq:v2},\ref{eq:v1} lead to a very different
relation between $V_2$ and $V_1$, namely $V_1 V_2^2 = V_0^3$. Thus in this regime, 
$V_1$ and $V_2$ renormalize in opposite ways. So if $V_1$ diverges, $V_2$ vanishes and vice-versa.
For $U_{fc}=0$, we can solve equation~\ref{eq:v2} to get 
\begin{equation}
V_2=V_0-\frac{\rho_0 V_0^3}{\epsilon_d} \ln\frac{D}{D_0}\,.
\end{equation}
Since $\epsilon_d < 0$, $V_2$ decreases logarithmically as $D$ decreases. As discussed above, $V_1$
increases concomitantly. Eventually, $V_2\rightarrow 0$ as $D\rightarrow T_K$, and hence $V_1$ diverges.
As is well known, this divergence is inherent in perturbative renormalization, and yields a closed form for
the Kondo scale as 
\begin{equation}
T_K\simeq D_0\exp\left(\frac{-|\epsilon_d|}{\rho_0 V_0^2}\right)\,.
\end{equation}
So, for $U_{fc}=0$, the standard expression for Kondo scale is reproduced~\cite{Hewson,Logan} and hence we establish that we can calculate Kondo scale from the scaling equations for hybridization. 
For finite $U_{fc}$, we obtain Kondo scale through a numerical solution of equations ~\ref{eq:v2} and ~\ref{eq:v1} in the next subsection.
\subsection{Numerical solution}
We use the usual Euler discretization to solve equations ~\ref{eq:v2} and ~\ref{eq:v1}. Again, we restrict to flows
of hybridization close to the Fermi level, so $\epsilon_k=0$. It is important to note that $\epsilon_d$, which also
enters these equations,  is also
a function of $D$, but its flow is determined by a second order equation. So, we compute the flow of $\epsilon_d$ first keeping 
$V_1=V_2=V_0$ constant, and use this flow to solve for the flow of $V_1$ and $V_2$. Another issue is the
discretization of the bandwidth. Since the Kondo scale is an exponentially small scale, the $D\rightarrow 0$
 regime must be sampled densely, while the non-universal scales can be sampled coarsely. Thus, a highly non-uniform
 grid has been used to discretize the conduction band, which gets progressively dense as $D\rightarrow 0$. 

\begin{figure}
\centering
\includegraphics[clip,scale=0.4]{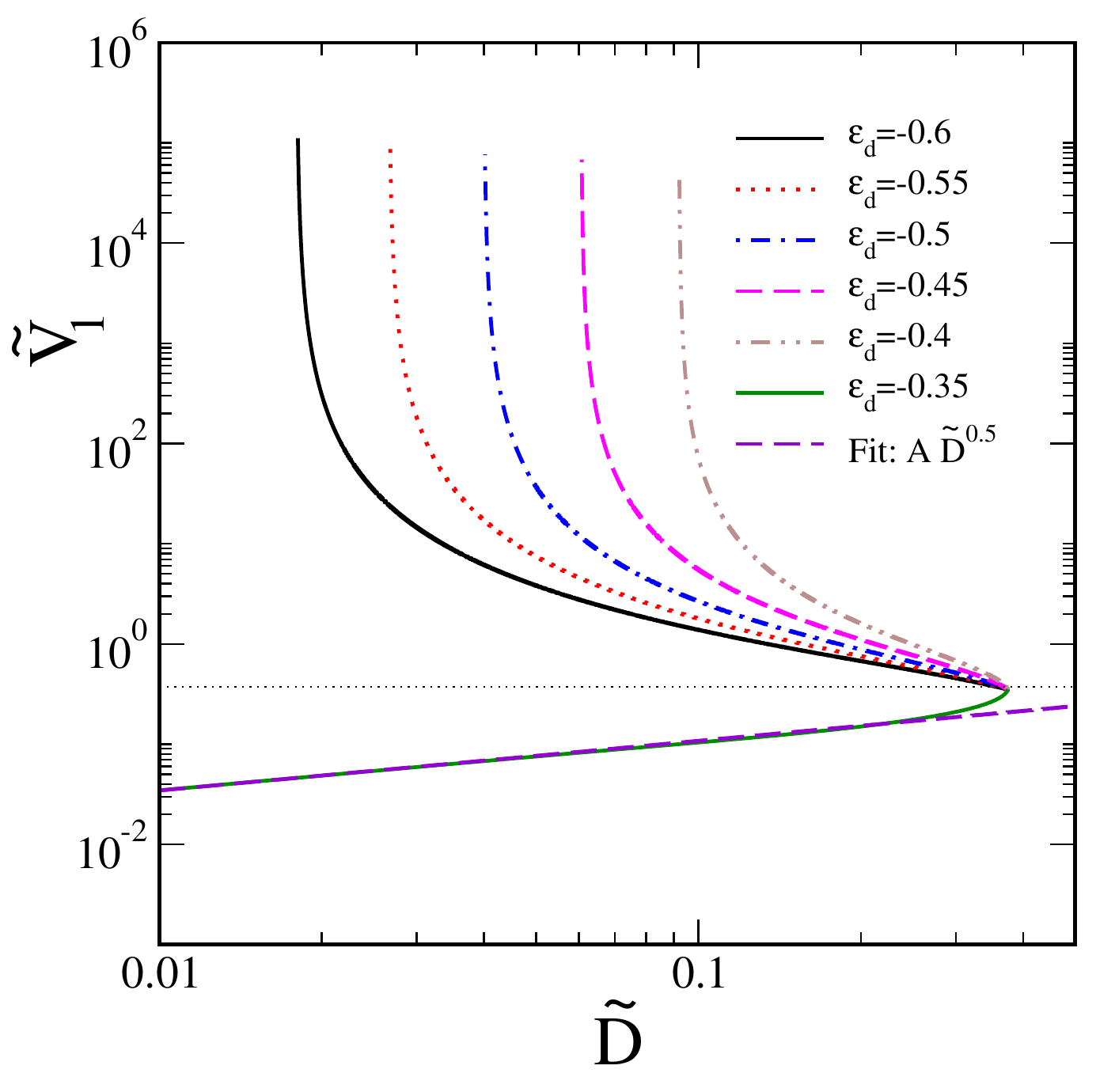}
\includegraphics[clip,scale=0.4]{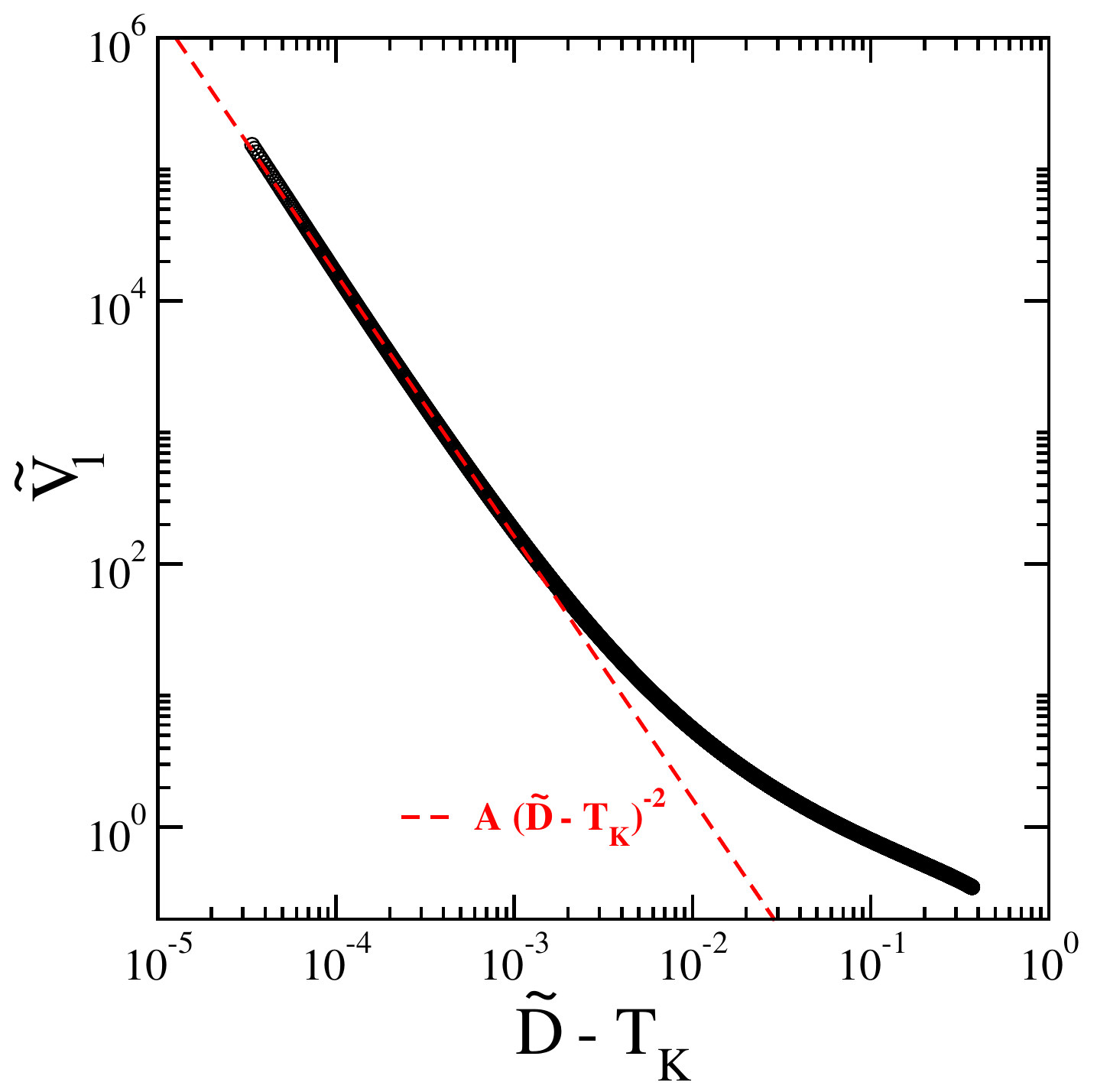}
\caption{Top panel: Scaling flow of $V_1$ for various values of $\epsilon_d$ with an initial value of $V_0=0.35$ and
the bare bandwidth, $D=0.375$.  Bottom panel: The divergence of $V_1$ shown in the top panel is analysed
 and it is found to be
of the form $(D-T_K)^{-2}$. Initial value of $\epsilon_{d}=-0.8$ and the Kondo scale $T_{K}\simeq 0.004$. }
\label{flow_V_1_var_ed}
\end{figure}

We fix the initial $V_1=V_2=0.35$, and investigate the effect of varying the initial $\epsilon_d$ on the flow
of hybridization in the top panel of figure~\ref{flow_V_1_var_ed}. For $|\epsilon_d| > D$, the $V_1$ diverges at a finite scale.
As discussed in section~\ref{sec:rehyb}, this scale is indeed the Kondo scale.
For $|\epsilon_d| < D$, the hybridization vanishes algebraically, and as we know from section~\ref{sec:reed}, the impurity
orbital energy also vanishes as $D\rightarrow 0$, thus implying a flow to the empty orbital regime.
It is interesting to note that in the flow to the empty orbital, the hybridization vanishes as $D^{1/2}$, while
the divergence in the LM regime is $\sim (D-T_K)^{-2}$ as shown in the bottom panel of figure~\ref{flow_V_1_var_ed}.

\begin{figure}
\centering
\includegraphics[clip=,scale=0.4]{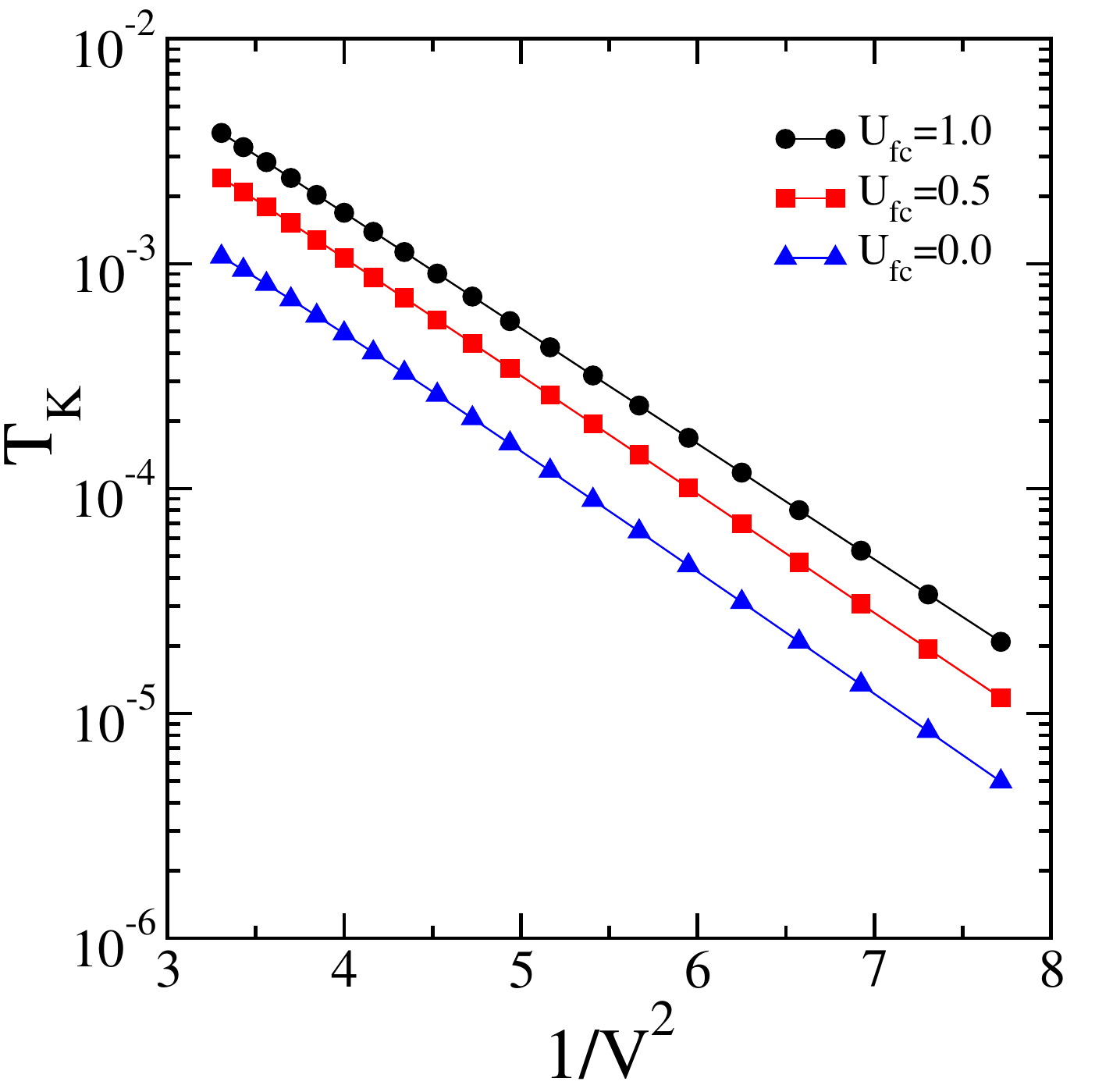}
\includegraphics[clip=,scale=0.4]{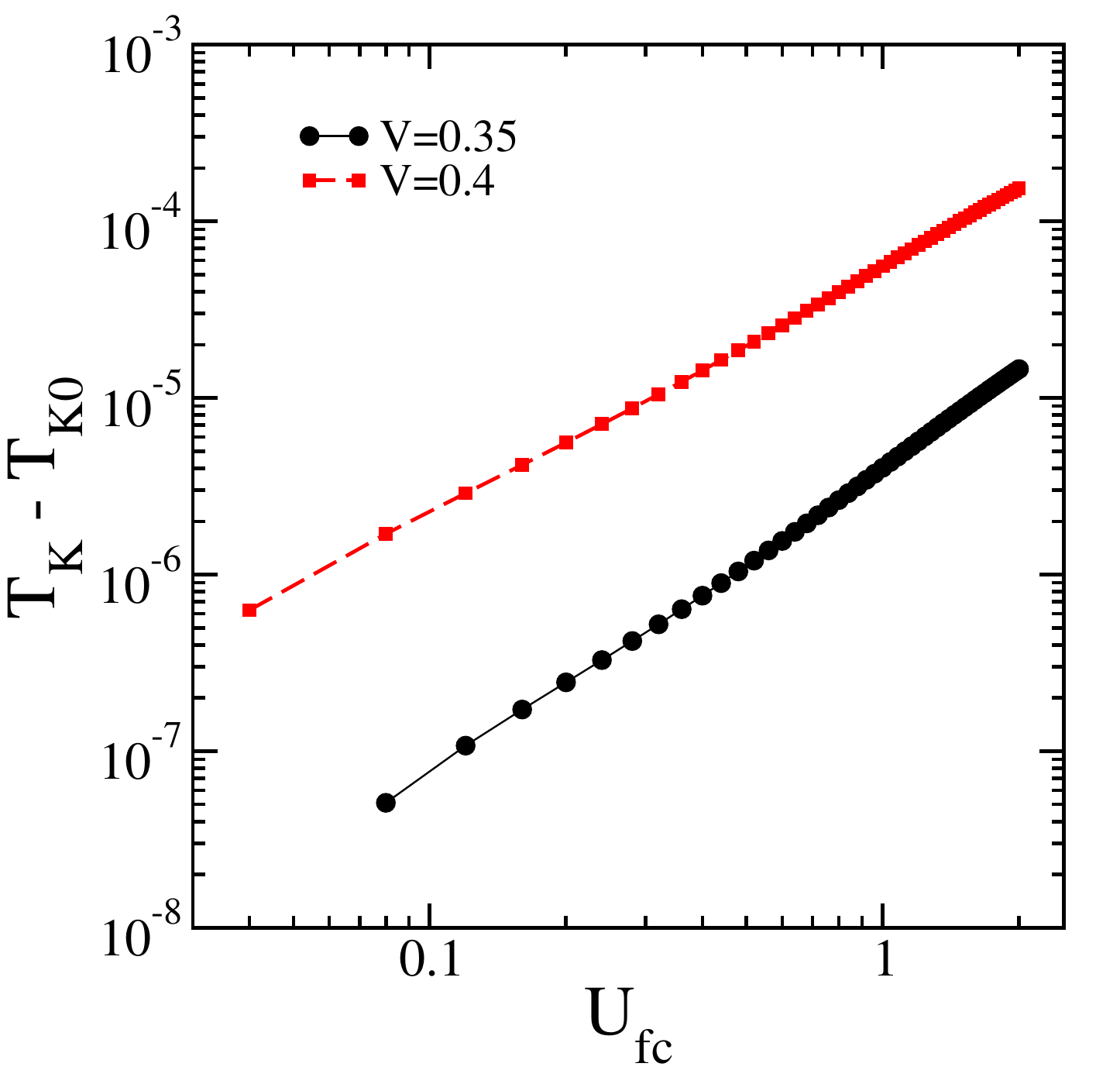}
\caption{Top panel: Kondo scale as a function of $1/V^2$ for three $U_{fc}$ values. Initial values of $\epsilon_{d}$ and D are -2.4 and 0.375 respectively.  Bottom panel: Kondo scale
as a function of $U_{fc}$ for two $V$ values showing a power law dependence of the shift in the Kondo scale. The exponents for $V=0.35$ and $0.4$ are $\sim 1.8$ and $\sim 1.4$      respectively. }
\label{kondo_scale}
\end{figure}

\begin{figure}
\centering
\includegraphics[clip,scale=0.4]{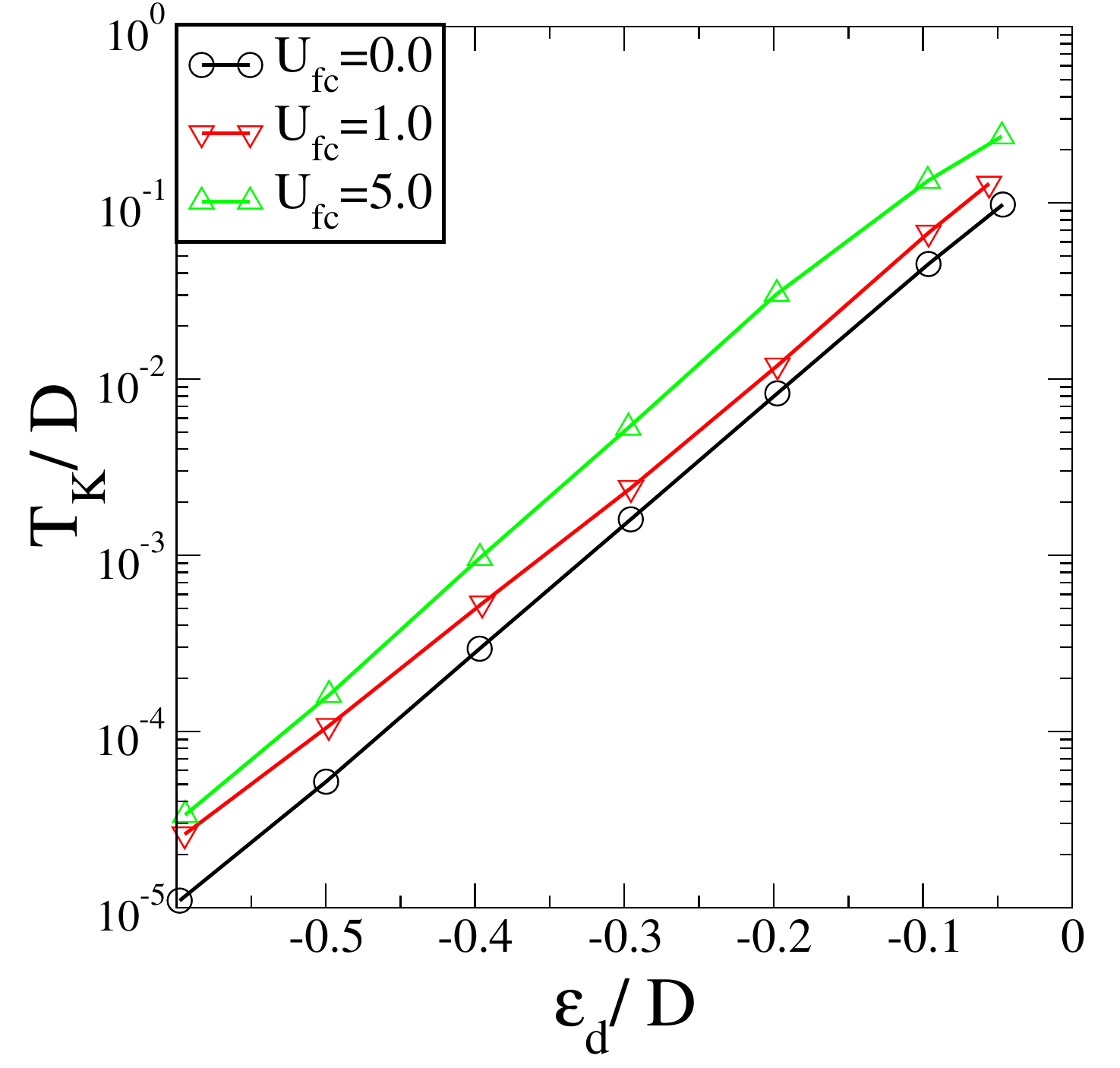}
\includegraphics[clip,scale=0.4]{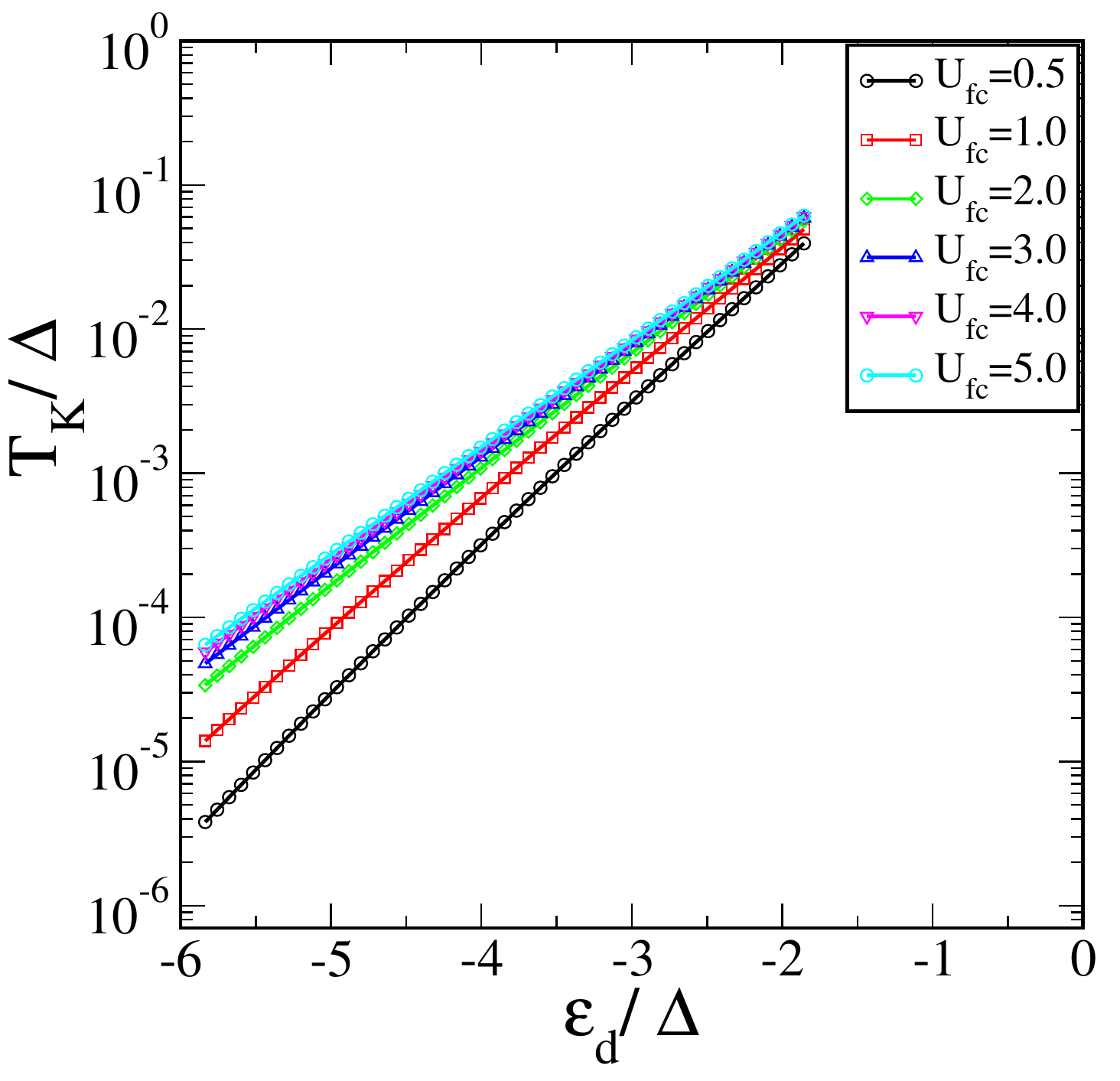}
\caption{Kondo scale as function of $\epsilon_{d}$ for increasing values of $U_{fc}$. Top panel shows the NRG data from figure 6(a) of Ref.~\onlinecite{Ryu}. The bottom panel is based on a numerical solution of the scaling equations~\ref{eq:v2},\ref{eq:v1} for hybridization. In both the figures, $T_{K}$ and $\epsilon_{d}$ are scaled with $\Delta$.}
\label{ed_TK_UFC}
\end{figure}
We can now investigate the effect of $U_{fc}$ on the Kondo scale in the regime of $|\epsilon_d| > D$. 
We fix the initial  $\epsilon_d=-2.4$, and $D=0.375$, and find the Kondo scale as a function of several initial $V$ values ranging from $0.35$ to $0.55$. The results presented in the left panel of figure~\ref{kondo_scale} show the Kondo scale ($T_K$) as a function
of $1/V^2$ for three $U_{fc}$ values of $0$ (black circles), $0.5$ (red squares) and $1.0$ (triangles). The linear dependence
of $T_K$ on $1/V^2$ when plotted on a linear-log scale indicates an exponential dependence, and indeed, the slope
correlates with $\epsilon_d$ (as shown in bottom panel of figure~\ref{ed_TK_UFC}). Again, the slope  does not depend strongly on  $U_{fc} $ indicating that it does
not enter the exponential, but in the prefactor. However, it is clear that the Kondo scale increases with increasing $U_{fc}$.
This is shown in the bottom panel of figure~\ref{kondo_scale}, where $T_K-T_{K0}$ (with $T_{K0}$ being the $U_{fc}=0$
scale) is seen to depend on $U_{fc}$ as a power law, i.e $T_K = T_{K0} + A U_{fc}^\gamma$. The exponent $\gamma$ is found to be $\sim 1.8$ and $\sim 1.4$ for $V=0.35$ and $0.4$ respectively, and hence decreases
with increasing $V$.   The  $U_{fc}$ driven upward renormalization of the Kondo scale maybe interpreted as a crossover
to a weaker coupling regime, and hence increased charge/valence fluctuations. The enhancement of Kondo scale with $U_{fc}$ was also reported in the NRG calculations~\cite{Hewson1,Irkhin2,Ryu}. Our results confirm this enhancement of Kondo scale with valence fluctuations. As shown above, we have also calculated the functional form of this dependence. In figure~\ref{ed_TK_UFC} Kondo scale is plotted as a function of impurity energy level for increasing values of $U_{fc}$. We have benchmarked our results with NRG, for which the data has been extracted from Ref.~\onlinecite{Ryu}.  As shown in figure~\ref{ed_TK_UFC}, Kondo scale depends exponentially on $\epsilon_{d}$ and gets enhanced with $U_{fc}$ which is what we have already shown in figure~\ref{kondo_scale}. Kondo scale depends on $\epsilon_{d}$ as $T_{K}\sim e^{\beta \frac{\epsilon_{d}}{\Delta}}$, so we extracted $\beta$ exponent for the case of $U_{fc}/\Delta=5$ and found it to be $\sim 1.62$ and $\sim 1.71$ from NRG data and our data respectively. Thus, we find excellent agreement of our results based on perturbative scaling with NRG calculations.

\section{Discussion}
In this section, we give a brief discussion of our results and compare our results with earlier studies of e-SIAM. We have solved the scaling equations for $\epsilon_{d}$ and hybridization analytically for tractable parameter regimes and solved them numerically as well. Our numerical solution does not rely on any special limits which are usually used to get the analytical solution of the scaling equations~\cite{Haldane,Jefferson}. Similarly, unlike what is usually done, we have not restricted to infinite U limit. Rather, we have explored the finite U case as well. Scaling equations have been solved both for particle-hole symmetric and asymmetric cases. In both cases, we find that scaling behaviour of the parameters depends on the initial value of the bandwidth. We observe different scaling flows for the cases when bandwidth D is greater and smaller than $\epsilon_{d}$.
For the particle-hole symmetric case, $\epsilon_{d}$ does not change under flow while $U$ scales upwards or downwards depending on whether  $D$ is smaller or greater than $\epsilon_{d}$. In presence of 
$U_{fc}$, the Hubbard repulsion $U$ scales down to weaker values for both the cases. Valence fluctuations are expected to be stronger in the particle-hole asymmetric case and hence $U_{fc}$ interaction is supposed to have stronger renormalization effects in this case. Our results confirm that in the presence of $U_{fc}$, the decrease in $U$ gets steeper. For $\epsilon_{d}$, we find that it scales upwards or downwards depending on whether $D > -\epsilon_{d}$ or $D < -\epsilon_{d}$. $U_{fc}$ interaction has enhanced the effect on this behaviour. However, as expected, it is in the mixed valent case that $U_{fc}$ has a stronger effect on the scaling flow of $\epsilon_{d}$.  One more interesting aspect about the scaling flows in the particle-hole asymmetric case is that they are non-monotonic unlike the ones in the symmetric case.
We have also calculated the scaling invariant for e-SIAM, and we find that it is quite different from that of SIAM which signifies that scaling flow of $\epsilon_{d}$ is not same in both models.

From the third order renormalization calculation of e-SIAM, we have found the scaling equation for hybridization which has been solved, both analytically, in suitable limits and numerically, for the general case. From the analytical solution of these scaling equations, we have calculated new scaling invariants of our model. Though scaling equation for hybridization was studied in Ref.~\onlinecite{Jefferson}, the scaling invariant was not explicitly derived. So we have calculated this scaling invariant both for $U_{fc}=0$(SIAM) as well as for e-SIAM, and once again we find that scaling invariants for these two models are not same which shows that $U_{fc}$ interaction changes the scaling trajectories of the Anderson impurity model parameters. One of the main results of our paper is the calculation of the functional dependence of Kondo scale on $U_{fc}$. Though it was already found that Kondo scale gets enhanced in the presence of $U_{fc}$ interaction,  we have found that there is power law dependence of Kondo scale on $U_{fc}$ interaction.

In NRG calculations of e-SIAM, it was  found that $U_{fc}$ interaction renormalizes the model parameters and consequently enhances the Kondo scale. In Refs.~\onlinecite{Hewson1,Hewson2}, the authors had found that $U_{fc}$ interaction does not change the strong coupling fixed point which is what we have also found that Kondo scale does get renormalized due to $U_{fc}$ but the contribution is only to the prefactor of the expression of the Kondo scale. 
However,  the numerical solutions of the scaling equations show clearly that flow of the parameters of the model is from Kondo regime to mixed valent regime and hence there is enhancement of valence fluctuations due to $U_{fc}$ interaction as has been also found by
NRG results from Katsnelson's group\cite{Irkhin1,Irkhin2}. We could not find that $U_{fc}$ makes Kondo fixed point unstable and hence within the validity of the perturbative renormalization method, we confirm that strong coupling point of e-SIAM is same as that of single impurity Anderson model.

We have benchmarked our results with NRG calculations and have found excellent agreement which signifies that our method has been able to calculate the renormalization effects of valence fluctuations in e-SIAM consistently. However,it needs to be noted that due to the perturbative nature of our renormalization method,we have not been able to explore the strong coupling fixed point of the model and hence could not capture the co-existence of the spin and charge Kondo effects which were found at the effective Hamiltonian level and have been confirmed by NRG calculation as well. We leave the full exploration of the strong coupling physics of the model for our future studies.

\section{Summary}
In this paper, we have investigated the effect of a repulsive interaction between the correlated impurity electrons and the
non-interacting conduction electrons in the extended single impurity Anderson model(e-SIAM) through unitary transformations
and perturbative renormalization of the model. A Schrieffer-Wolff transformation of the e-SIAM shows that the strong coupling regime is governed by a spin-charge Kondo model, unlike the Anderson impurity model where spin fluctuations dominate the strong coupling physics. Through perturbative renormalization to second and third order (following Haldane and Jefferson respectively), we found the scaling equations of the model parameters, with a focus on the effect of $U_{fc}$ on the renormalization flows. The scaling invariants of the model were also found. A divergence in the flow of hybridization, signaling the breakdown of perturbative renormalization,
 can be used to identify a low energy scale, and is shown to be the Kondo scale, through analytical arguments and a numerical
 solution of the scaling flows. The $U_{fc}$ interaction leads to an increase in the Kondo scale through a renormalization
 of the prefactor, and hence may be interpreted as leading to enhanced valence fluctuations. Our results are in agreement with earlier NRG studies where it was found that $U_{fc}$ leads to the renormalization of the Anderson model parameters. We have also confirmed that Kondo scale gets enhanced due to $U_{fc}$ interaction which was found in NRG calculation.From the numerical solution of the scaling equations of e-SIAM, we find that the system flows to the mixed valent regime. However, based on our perturbative renormalization method we were not able to show conclusively that Kondo fixed point becomes unstable to $U_{fc}$ interaction. In future, we would like to do flow equation renormalization study of e-SIAM to explore the regimes when $U_{fc}$ interaction becomes stronger or comparable to Hubbard repulsion.

\begin{acknowledgments} 
We acknowledge discussions with Subroto Mukerjee, Jozef Spalek and 
Arghya Taraphder. We also acknowledge funding from JNCASR and DST, India. 
\end{acknowledgments}

\end{document}